\documentclass[aps,twocolumn,prl,showpacs,color,psfig,epsf]{revtex4-1}
\usepackage{amsmath}
\usepackage{graphicx}
\usepackage{dcolumn}
\usepackage{bm}
\usepackage{amssymb}
\usepackage{multirow}
\usepackage{color}
\usepackage{hyperref}
\usepackage{float}
\usepackage{color}
\usepackage{amsfonts}
\usepackage{epsf}
\usepackage{graphicx}
\usepackage{natbib}

\definecolor{red}{rgb}{1,0,0}
\definecolor{blue}{rgb}{0,0,1}
\definecolor{darkgreen}{RGB}{3,101,3}
\definecolor{antonio}{RGB}{51,153,255}
\definecolor{francesco}{RGB}{255,0,255}

\begin{document}

\title{Work fluctuations of self-propelled particles in the phase separated state}

\author{P. Chiarantoni $^1$, F. Cagnetta $^2$, F. Corberi$^3$,  G. Gonnella$^1$, A. Suma$^{1,4}$}
\affiliation{$^1$ Dipartimento di Fisica, Universit\`{a}  di Bari, {\rm and} Sezione INFN di Bari, via Amendola 173, 70126 Bari, Italy}
\affiliation{$^2$ SUPA, School of Physics and Astronomy, University of Edinburgh, Edinburgh EH9 3FD, United Kingdom}
\affiliation{$^3$ Dipartimento di Fisica E. R. Caianiello {\rm and} INFN, Gruppo Collegato di Salerno, Universit\'{a} di Salerno, via Giovanni Paolo II 132, 8408 Fisciano (SA), Italy}
\affiliation{$^4$ Institute for Computational Molecular Science and Department of Biology, Temple University, Philadelphia, PA 19122, USA}

\begin{abstract}

We study the large deviations of the distribution $P(W_\tau)$ of the work associated with the propulsion of individual active brownian particles in a time interval $\tau$, in the region of the phase diagram where macroscopic phase separation takes place. $P (W_\tau )$ is characterised by two peaks, associated to particles in the gaseous and in the clusterised phases, and two separate non-convex branches. Accordingly, the generating function of $W_\tau$'s cumulants displays a double singularity. We discuss the origin of such non-convex branches in terms of the peculiar dynamics of the system phases, and the relation between the observation time $\tau$ and the typical persistence times of the particles in the two phases.

\end{abstract}

\maketitle


\section{Introduction}

The physical observables of equilibrium systems sit at the minima of thermodynamic potentials with small fluctuations regulated by Boltzmann-Einstein expressions~\cite{landau:5}. A general  framework for the description of larger fluctuations, holding also beyond equilibrium and static phenomena, is given by the theory of large deviations~\cite{touchette2009aa}. Consider a quantity $W_\tau =\sum _{i=1}^\tau W_i$, namely the sum of a large number $\tau$ of stochastic variables: if  a {\it large deviation principle} (LDP) holds, the asymptotics of the probability distribution $P(W_\tau)$ is characterized by a $\tau $-independent \emph{rate function} $I(w)$,
$w=W_\tau/\tau$ being the empirical average, such that
\begin{equation}\label{eq:LDP} -\lim_{\tau\rightarrow\infty} \frac{1}{\tau}\ln{\left\lbrace P\left( W_{\tau}\right)\right\rbrace} = I(w). \end{equation}
In equilibrium statistical mechanics,
considering for instance $W$ as the energy, and $\tau$ the number of particles,
Einstein's theory of fluctuations states that the entropy is the rate function,
while its Legendre-Fenchel transform is the free energy.
When the former is convex, the latter is a regular function.
Conversely, non-convexity implies a singular free energy and
the lack of ensemble equivalence, occurring, for instance,
in the presence of first
order phase-transitions.

Via large deviation theory, one can define the analogue of entropy and free-energy functions for dynamical problems, where $\tau$ is an observation time and $W_\tau$ a time-additive functional of the system trajectory in $[0,\tau]$ (such as the entropy production or the work done by some force~\cite{TOUCHETTE20185}).
A trajectory-based description~\cite{ruelle2004thermodynamic,lecomte2007thermodynamic} has been adopted for both general considerations, such as proofs of the fluctuation theorem~\cite{gallavotti1995dynamical,kurchan1998fluctuation,lebowitz1999gallavotti}, and model-specific studies, in the context of sheared fluids~\cite{evans2004rules} or in that of the glass transition~\cite{Merolle10837,garrahan2007dynamical,chandler2010dynamics}. The goal of this letter is to use this description to characterise fluctuations in systems of active particles undergoing motility-induced phase separation (MIPS).

In active matter systems, the basic units consume internal energy resources to establish a permanent non-equilibrium condition where work is done on the surrounding environment. This work, which will be later defined for our model 
of self propelled particles (SPP), is the observable $W_{\tau}$ considered in this paper.  The interest in systems of SPPs was fostered by the display of properties without any analogue in passive systems~\cite{Marchetti13,Elgeti15,gonn15,carenza2019}, including MIPS~\cite{Tailleur08,Ginelli10,Peruani06,fily2012athermal,Redner13,Stenhammer13,Buttinoni13b,theers2018clustering,Suma13,Suma14,cugliandolo2017phase,digregorio2018full}, spontaneous alignment~\cite{caprini2020spontaneous}, and other relevant properties such as accumulation at the system boundaries~\cite{Elgeti09,bechinger2016active,Elgeti13,costanzo2012transport,yang2014aggregation,caprini2018active}, or the possibility to build bio-driven microgears~\cite{angelani2009self,pietzonka2019autonomous}. Besides, the presence of an internal mechanism with its own time- and length-scales, as is self-propulsion, causes also fluctuations to display nontrivial features~\cite{gradenigo2013aa,seifert2016newjPhy,whitelam2018phase,NemotoCates2019,tociu2019dissipation,fodor2020dissipation,Gradenigo_2019,mallmin2019comparison,cagnetta2019emergent}.

Understanding such properties is instrumental to the development of a stochastic thermodynamics of active
suspensions, a topic which has attracted much attention in recent times~\cite{seifert2011stochastic,ganguly2013stochastic,chaudhuri2014active,kumar2015aa,shankar2018hidden,chaki2018entropy,chaki2019effects,dabelow2019irreversibility,caprini2019entropy}.

In this context, we have examined in~\cite{cagnetta2017large} the large deviations of the individual work $W_{\tau}$ done by the self-propulsion force which pushes one active particle, both in the dilute limit and at small but finite density. Here we repeat the analysis for denser suspensions, for which MIPS occurs.
We find a double-peaked probability distribution, so that the corresponding
rate function of Eq.~(\ref{rateft})
presents two distinct non-convex branches---a feature not typical of the thermodynamic potentials of 
passive phase coexistence~\cite{touchette2009aa}. 
We show that different sectors of the rate function correspond to dynamical trajectories of individual particles with different
qualitative features. This sheds a new light on the MIPS transition and
reveals a rich dynamical behaviour of the cluster phase, complementing the results of~\cite{NemotoCates2019} where the collective work done by \emph{all} the active forces in the system was studied.

\section{The model}

We consider, as customary~\cite{fily2012athermal}, a system of $N$ self propelled disks of diameter $\sigma_d$, 
with only soft excluded volume interaction, in a two-dimensional square box of side $L$ with periodic boundary conditions. Each particle is propelled by an active force with modulus $F_{\rm act}$ and direction $\bold{n}_i=( \cos{\theta_i(t)},\sin{\theta_i(t)})$. The $i$-th particle position $\bm{r}_i$ and orientation $\theta_i$ obey 
\begin{equation} \label{eq:langevin} m\ddot{\bold{r}}_i=-\gamma\dot{\bold{r}}_i+F_{\rm act} \bold{n}_i- {\boldsymbol{\nabla}}_i\sum_{j(\neq i)}U(r_{ij}) + \bm{\xi}_i \; ,\quad \\ \dot{\theta}_i=\eta_i \; , 
\end{equation} 
where $r_{ij}=|{\bold{r}}_i-{\bold{r}}_j|$ is the inter-particle distance and $U(r)$ is a purely repulsive potential 
$U(r)=4\varepsilon [({\sigma}/{r})^{64}-({\sigma}/{r})^{32}]+\varepsilon$ if 
$r< 2^{1/32}\sigma$ and $0$ otherwise, 
$\sigma_d=2^{1/32}\sigma$ in order to have the potential truncated at its 
minimum, set equal to the disks diameter~\cite{cugliandolo2017phase,digregorio2018full}.
$\bm{\xi}$ and $\eta$ are zero-mean Gaussian noises satisfying $\langle \bm{\xi}_{i}(t) \, \bm{\xi}_{j}(t') \rangle = 2 \gamma k_B T \delta_{ij} \delta(t-t') \bold{1}$ and $\langle \eta_{i}(t) \, \eta_{j}(t') \rangle = 2 D_{\theta} \delta_{ij} \delta(t-t')$. The units of length, mass and energy are given by $\sigma_d$, $m$ and $\varepsilon$, respectively, and are set to one. The rotational diffusion coefficient is set to $D_\theta = 3\gamma k_BT/{\sigma^2_d}$~\cite{fily2012athermal}. The controlling parameters are the packing fraction $\phi =\pi{\sigma^2_d}N/(4L^2)$, which is tuned fixing $N$ and varying $L$, and the P\'eclet number $Pe = F_{\rm act} {\sigma_d}/(k_BT)$, which we change by varying $F_{\rm act}$ at $\gamma=10$ and $k_BT=0.05$. For this choice of parameters, inertial terms are typically negligible. The phase diagram of this system has been studied in detail~\cite{digregorio2018full,klamser2018thermodynamic,levis2017}. For $Pe \ge Pe_c \approx 38$, when  $\phi$ exceeds a P\'eclet dependent threshold, an initial homogeneous state separates into a dense and a gaseous phase. 
\begin{figure}[!h]
\begin{center}
  \begin{tabular}{c}
       \includegraphics[width=1\columnwidth]{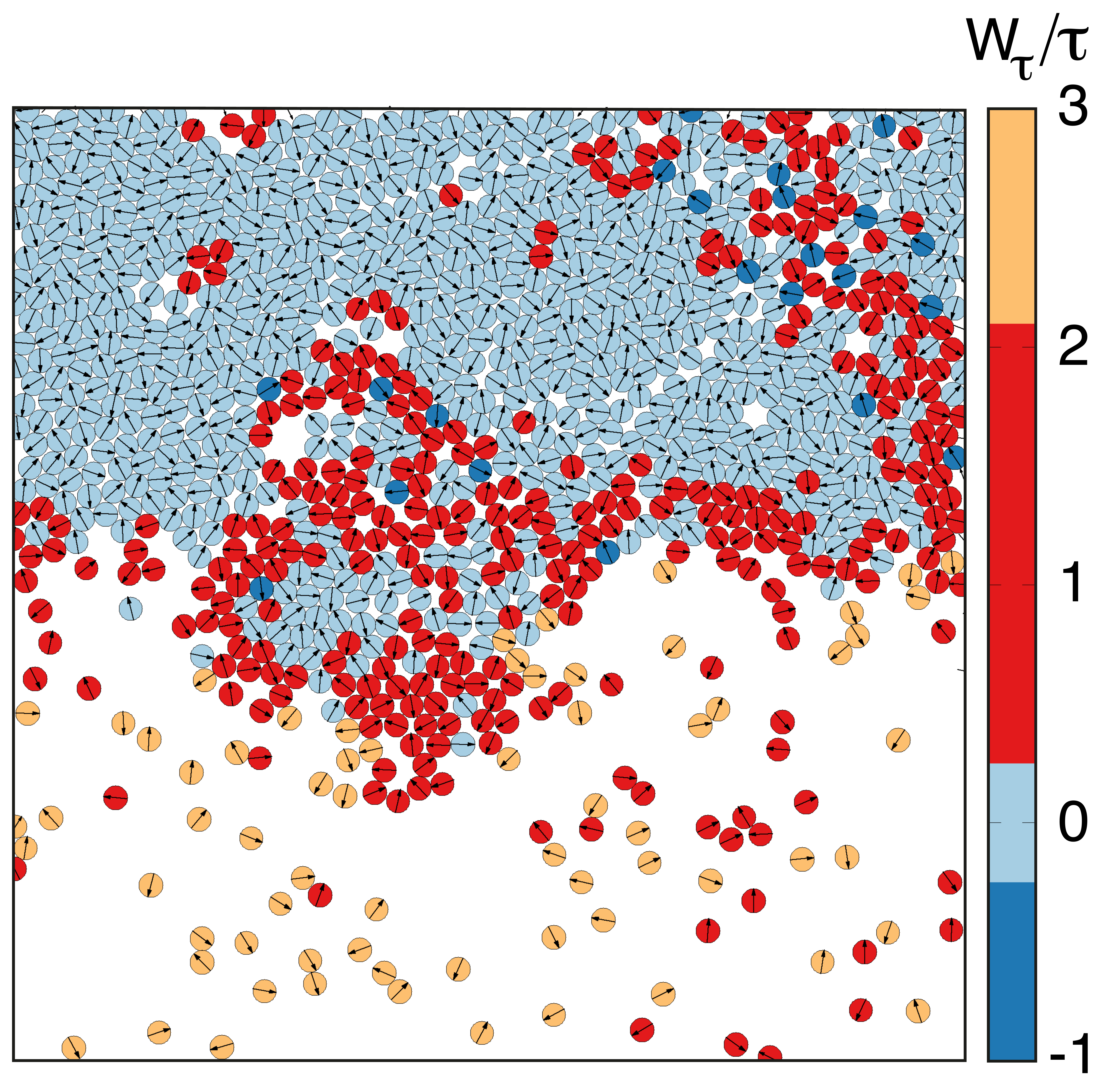}\\
    \end{tabular}
  \caption{Typical snapshot of a portion of the system at the interface between the coexisting
    dilute and dense phases at $Pe = 100$,  $\phi=0.5$ and $N=256^2$. Particles are colored according to $W_{\tau}/\tau$, with $\tau=10$. }
\label{fig1}
\end{center}
\end{figure}

A typical configuration of the system in the phase separated region  $Pe \ge Pe_c$ is shown in Fig.~\ref{fig1}. One observes a large aggregate of particles coexisting with a gaseous phase: due to activity, small clusters roam the gaseous phase, while  the dense phase is filled with holes~\cite{tjhung2018cluster} and other 
defects~\cite{digregorio2019clustering}.

\section{Active work}

We evaluate, for each disk $i$, the {\it active work} \begin{equation}\label{defw} W_{\tau} =  F_{act}\int_0^{\tau}{ \bold n_i(t) \cdot \dot{\bm{r}}_i(t)\, dt},\end{equation} where $t=0$ corresponds to some time after the system has reached a stationary state. $W_{\tau}$ represents the steady-state, single-particle contribution to the entropy production~\cite{kurchan1998fluctuation}, and, in general, it has been proposed as a relevant observable for the thermodynamical description of phase transitions in ABP systems~\cite{krinninger2016nonequilibrium,krinninger2019power}. The probability distribution $P(W_\tau )$ is the object of our investigation. The quantity 
\begin{equation}\label{rateft} 
I_{\tau}(w) = -\frac{1}{\tau}\ln \left.P(W_{\tau})\right|_{W_{\tau}=w \tau}
\end{equation} 
yields the rate function in the large $\tau$ limit, i.e. $I(w)=\lim_{\tau\rightarrow\infty} I_{\tau}(w)$.

\begin{figure}[!h]
\begin{center}
  \begin{tabular}{c}
       \includegraphics[width=1\columnwidth]{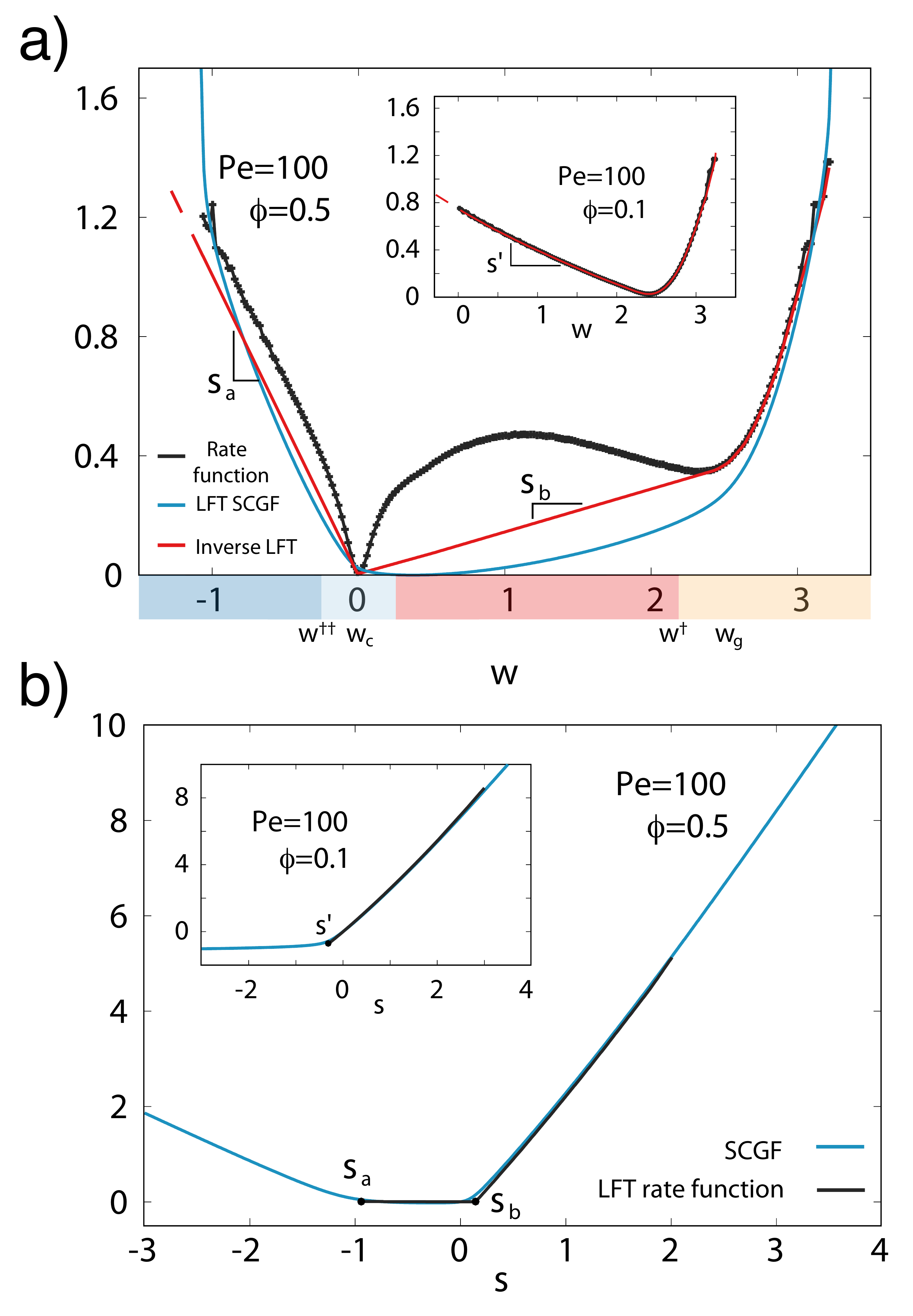}\\
    \end{tabular}
  \caption{a) Rate function at $Pe = 100$, $\phi=0.5$, $N=256^2$ (black), for $\tau=10$.
    The blue curve is the LFT of the SCGF (shown in Fig.~\ref{fig2}b as a blue line), 
    for the same choice of $Pe$ and $\phi$.
    The red curve is the inverse LFT of the LFT of $I_\tau (w)$ (black curve in Fig.~\ref{fig2}b). 
    All the functions in Fig.~\ref{fig2}a
    have been translated vertically to set the minimum in zero.
    The color bar corresponds to the one of  Fig.~\ref{fig1}. In the inset, the same quantities are plotted
    for $\phi =0.1$, where the system is homogeneous.
    b) Comparison between the SCGF (blue) and the LFT of $I_\tau (w)$ (black).
    Fluctuations with $s\gtrsim 2$ (corresponding to the maximum sampled slope of
    $I_\tau(w)$, occurring at $w\gtrsim 3$) are not observed in the
    LFT transform of $I_\tau$ due to limited statistics.
    In the inset, the same quantities are plotted
    for $\phi =0.1$.} 
\label{fig2}
\end{center}
\end{figure}
Fig.~\ref{fig2}a shows $I_{\tau}(w)$ at $Pe=100$ and $\phi=0.5$, as in Fig.~\ref{fig1}, while, in the inset, $\phi=0.1$ (for the same $Pe$), as in~\cite{cagnetta2017large}. Here $\tau=10$, but the structure of $I_{\tau}$ is preserved at least up to
$\tau\sim 500$~\footnote{Recall that the persistence time $D_{\theta}^{-1} = 0.67$ in our units: this is what $\tau$ shall be compared to}, as we will discuss further below. In the homogeneous state (for $\phi=0.1$, see inset), the rate function has a single minimum with a linear branch departing from its left, a fact that was interpreted as due to a condensation transition occurring at smaller-than-average $w$'s~\cite{cagnetta2017large}. In the MIPS region, instead, $I_\tau(w)$ shows two minima at $w=w_c$ and $w=w_g>w_c$, corresponding to the typical values of $w$ for particles belonging to the  cluster (light blue in Fig.~\ref{fig1}) and those in the gaseous phase (yellow in Fig.~\ref{fig1}). Therefore, this structure is a natural manifestation of the two phases coexisting in the system---one where particles are normally propelled by the active forces and another, jammed, where those forces do a smaller work due to steric hindrance.

Fig.~\ref{fig2}a also elucidates the different character of our approach with respect to that of~\cite{NemotoCates2019}, where $W_{\tau}$ is summed over all the SPPs.
In doing this sum the contributions of particles in the gaseous and dense phase are mixed. The distribution of the total active work is consequently peaked at the average between the typical work done by particles in the gas and the one done by those in the solid. By contrast, the distribution of the work done by individual particles, as we show in this paper, remains bimodal up to very long times. For a very large system, particles deep inside the dense phase have a very low chance of leaving the cluster. Correlations between the active work produced by these particles will then persist up to very long times, causing a significant difference between the statistics of the individual's active work and that of the sum.
However, the approach of~\cite{NemotoCates2019} is more suited to detect some other features of the system, such as collective motion.

Besides an overall similarity of our $I_\tau$ with usual thermodynamic potentials in first order phase transitions, some new features emerge, notably non-convexity on the left of $w_c$. The implications of non-convexity are better understood by resorting to the scaled cumulant generating function (SCGF),
\begin{equation} G_{\tau}(s)  = \frac{1}{\tau}\ln \left\langle e^{s W_{\tau}}\right\rangle. \label{defsgcf}
\end{equation}
The large-$\tau$  limit of the SCGF, $G(s)=\lim _{\tau\to\infty}G_\tau(s)$, coincides with the Legendre-Fenchel 
transform (LFT) of $I(w)$, $G(s)=\sup_w{\left\lbrace sw-I(w) \right\rbrace}$. When convex, also $I(w)$ is the LFT of $G(s)$. The two functions, in fact, can be regarded as thermodynamic potentials associated to a microcanonical (fixed $W_{\tau}$) and canonical ($W_{\tau}$ fixed on average by a bias $s$) ensembles of trajectories with a given $W_{\tau}$, respectively. $G_{\tau}(s)$ can be computed directly in our simulations and is shown in Fig.~\ref{fig2}b as a blue solid line; parameters are the same as in Fig.~\ref{fig2}a. Despite the finiteness of $\tau$, there is a good agreement between $G_{\tau}(s)$ and the LFT of $I_{\tau}(w)$, which is shown in the same figure as a black solid line. 
Let us then discuss how the different branches of $I_\tau(w)$ and $G_\tau(s)$ are mapped into each other, and
the nature of the associated particle trajectories.   


$I_\tau (w)$ is well approximated by a Gaussian for $w$ larger than a certain threshold value $w^\dag$. Comparing Fig.~\ref{fig2} with Fig.~\ref{fig1}, one concludes that this range corresponds to the yellow particles in the gaseous phase, for which many body effects can be neglected.
The corresponding branch of $G_{\tau}$, for $s>s_b\simeq 0.14$
(the slope of the convex envelope between the two minima of $I_\tau (w)$, red line in
Fig.~\ref{fig2}a), is quadratic in $s$, even though it might appear flat on the scales of Fig.~\ref{fig2}b. Proceeding towards lower $w$'s, the concave sector of $I_\tau(w)$ between the two minima is mapped by the LFT into the single point $s=s_b$, where the LFT displays a discontinuous first derivative. Typical trajectories contributing to this
sector are those of the red particles in Fig.~\ref{fig1}. 
Going further, the fluctuations described
by $I_\tau (w)$ around the minimum in $w_c$ are also approximatively Gaussian,
hence $G_\tau(s)$ is again parabolic in the corresponding sector $s_a<s<s_b$
(as for $s>s_b$, on the scale of the figure this part looks rather flat). The particles which contribute to this branch
are the light blue particles of Fig.~\ref{fig1}, which are stuck inside
the aggregate. For $w$ smaller than a certain value $w^{\dag \dag}$, $I_\tau(w)$ is
linear, thus its LFT diverges.
$G_{\tau}(s)$, instead, does not, due to the limited sampling. These fluctuations are originated by the dark blue particles in the cluster of Fig.~\ref{fig1}.

Notice that all the information relative to the trajectories of the red particles,
contained in the concave branch of $I_{\tau}$, is lost when going from
$I_\tau(w)$ to $G_\tau (s)$, due to the non-involutivity of the LFT
of concave functions. This property is strictly connected to statistical ensemble
inequivalence and phase transitions~\cite{ellis2007entropy,corberi2013aa,corberi2017development,corberi2015large,zannetti2014ab,zannetti2014aa}, which extends to the ensemble of trajectories considered
here~\cite{chetrite2013aa}. As a result, when a dynamical phase transition
occurs, the transform of $G_{\tau}$ does not yield back $I_{\tau}$, but only its convex envelope
(the red solid line in Fig.~\ref{fig2}a).

\begin{figure*}
       \includegraphics[width=2.1\columnwidth]{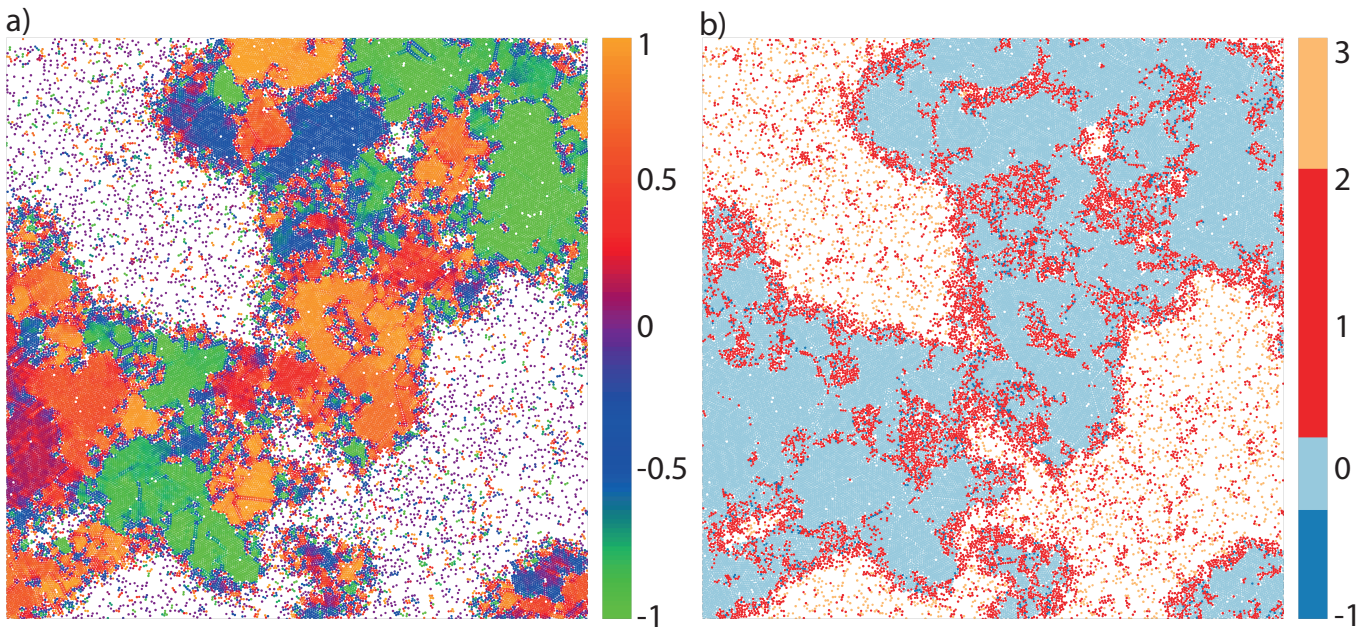}
  \caption{Snapshot of a portion of size $250\,\sigma_d$ of a system with $N=512^2$
    particles, at $Pe = 100$ and $\phi=0.5$. a) Particles are colored according to the
    hexatic parameter related to the local orientation of the triangular lattice
    occupied by the particles~\cite{Note2}.
    b) Same configuration as in a), but colored according to $W_{\tau}/\tau$,
    with $\tau=10$, as in Fig.~\ref{fig1}.} 
\label{fig3b}
\end{figure*}

\begin{figure}[!h]
\begin{center}
  \begin{tabular}{c}
       \includegraphics[width=0.9\columnwidth]{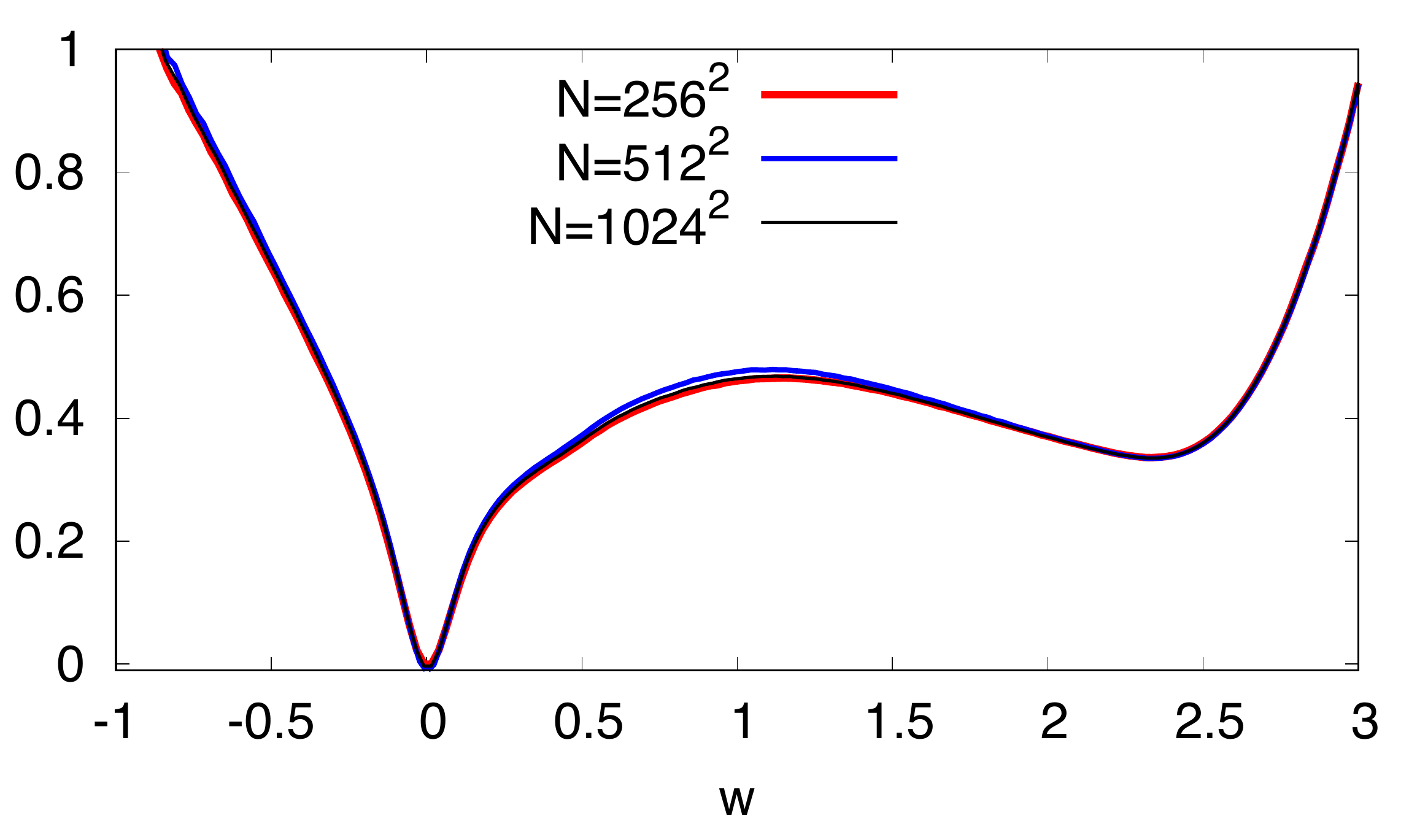}\\
    \end{tabular}
\caption{Rate functions at Pe$=100$,  $\phi=0.5$, $\tau=10$ and three 
values of $N$, see key. Similar results are found for different $\tau$ up to $\tau=500$. All the functions have been translated vertically to set the minimum in zero.} 
\label{fignew}
\end{center}
\end{figure}


\begin{figure*}
\begin{center}
       \includegraphics[width=2\columnwidth ]{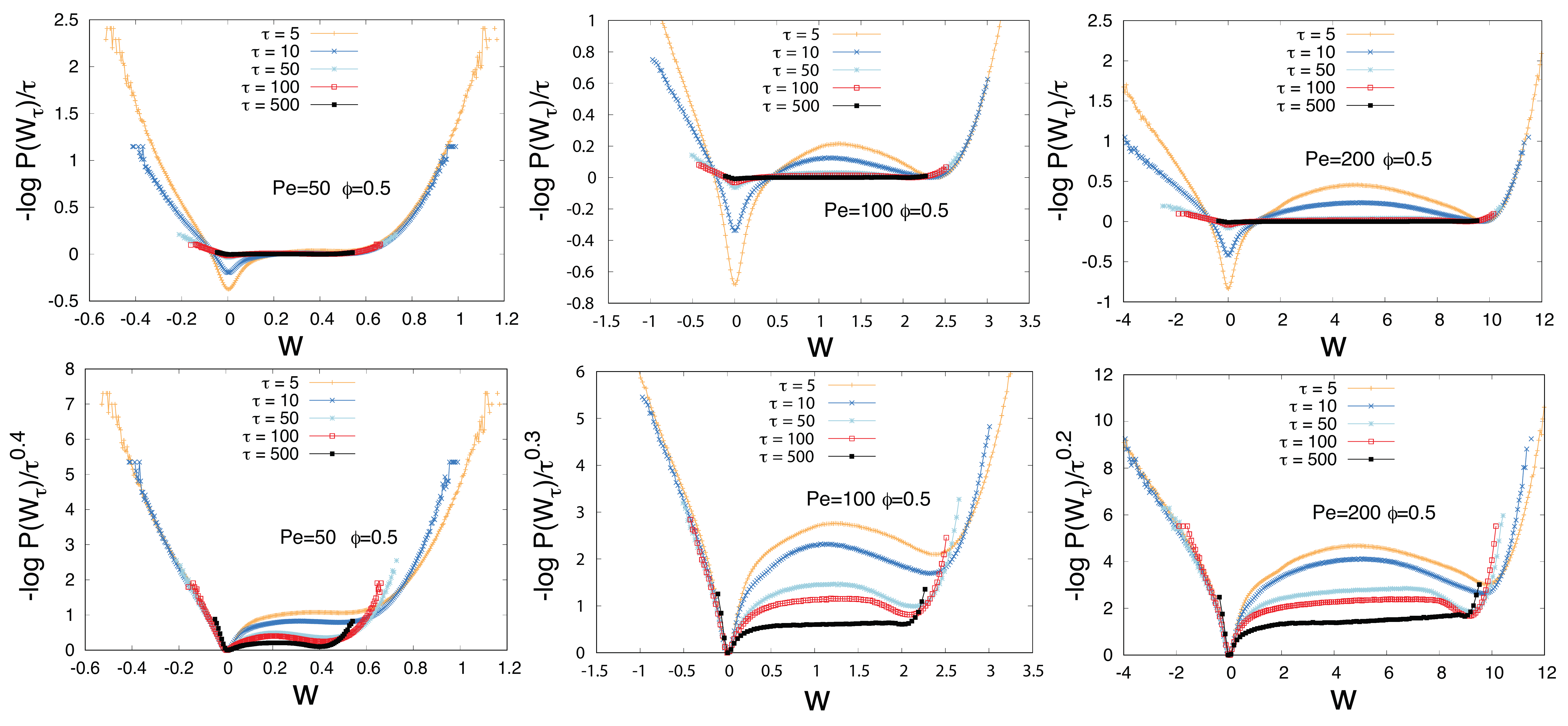}
       \caption{Behavior of the same rate function of Fig.~\ref{fig2} (for 
$Pe=50,100,200$ from left to right) as $\tau$ is changed. In the upper row $\tau^{-1}\ln P(W_\tau)$ is plotted against 
$w$, while in the lower row we plot $\tau^{\beta}\ln P(W_\tau)$, with $\beta=0.4, 0.3, 0.2$, respectively, for $Pe=50, 100, 200$. The functions of the first (second) row have been translated vertically to set the minimum in $w_g$ ($w_c$) to zero.} 
\label{fig3}
\end{center}
\end{figure*}

\subsection{Linear branches}

$I_\tau(w)$ is endowed with two linear branches, one to the immediate left of $w^\dag$
and the other for $w<w^{\dag \dag}$. The first is produced by 
those particles in the gaseous phase which, by hitting other disks or small
clusters, do a reduced work. This feature of $I_\tau(w)$, which can be thought of as a transition of fluctuations,  is also manifest at low densities, where MIPS
does not occur, as it has been reported in~\cite{cagnetta2017large}.
In fact, this linear branch is better observed in a dilute system,
as shown in the inset of Fig.~\ref{fig2}a, where it
extends for all $w<w^{\dag}$. In the presence of MIPS, instead, particles can do a reduced work by hitting the main cluster, thus the branch merges with the dense-phase minimum at $w_c$.

Let us now discuss the other linear part of $I_\tau(w)$, namely
that to the left of $w^{\dag \dag}$ which is originated inside the cluster and, therefore,
only present when MIPS occurs. As it can be seen in
Fig.~\ref{fig3b}, the particles inside the cluster which are able to move,
and hence to do some work, are concentrated along the boundaries of hexatically ordered domains, which are coloured differently
in Fig.~\ref{fig3b}a~\footnote{The hexatic parameter is a complex number defined here for each particle $k$ as $\psi_{k}=\sum_{j=0}^{N^k} \exp(6i\theta_{kj})/N^k$, where the sum over $j$ is for the $N^k$particles distant from $k$ less than a cutoff set to $1.1$, and $\theta_{kj}$ is the angle formed by the segment that connects the center of the $k$ and $j$ disks  between the two particles, using the x-axis as a reference. Particles in Fig.~\ref{fig3b}a are colored according to the projection between the hexatic parameter and the system's hexatic average value. Particles with the same local color have their hexatic projection ordered on the same direction~\cite{cugliandolo2017phase}.}. Most of these
particles are red in Fig.~\ref{fig3b}b, meaning that they contribute (together with other red particles
in the gaseous phase) to the region between the two minima of $I_\tau(w)$.
The particles which move inside the cluster can also push other disks
against their propulsion force: these back-pushed trajectories 
give rise to the branch with $w<w^{\dag \dag}$. This phenomenon occurs both along the boundary between the dilute and the dense phase and between different hexatic patches inside the aggregate, as shown in the two supplemental movies S1-S2 and in Fig. 3 of the SI.
In conclusion, both the linear branches of $I_\tau(w)$ are originated
by particles whose work is limited by the dragging due to the others.

\begin{figure}[!h]
\begin{center}
  \begin{tabular}{c}
       \includegraphics[width=0.9\columnwidth]{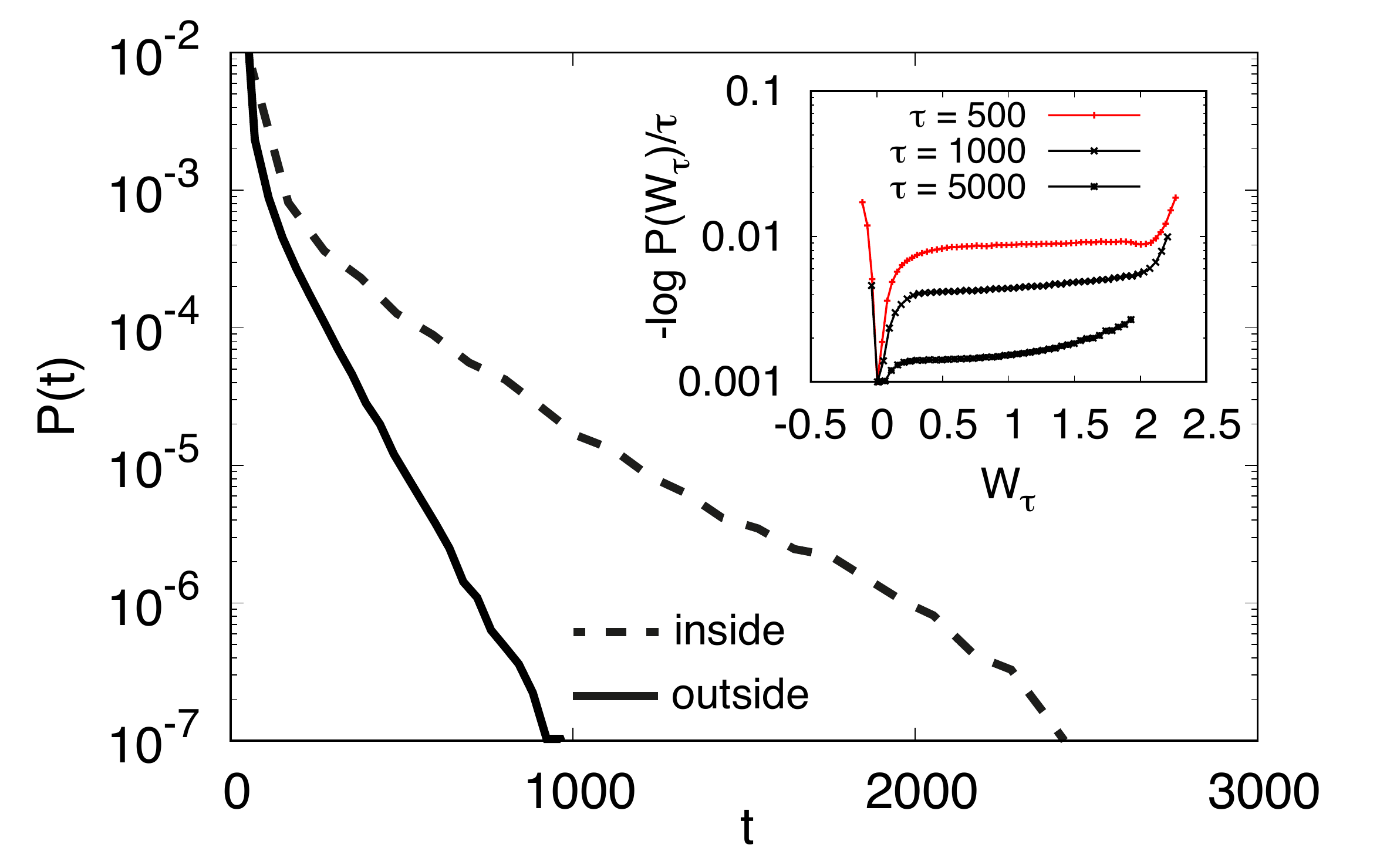}\\
    \end{tabular}
\caption{Distributions of times for particles persisting inside a cluster (dashed line) and outside the cluster (continuous line), for colloids at $\phi=0.5$ and $Pe=100$. A fit using $e^{-t/\tau_{res}}$ provides $\tau_{res}=265$ and 
$\tau_{res}= 92$ for particles inside and outside the cluster, respectively.  
Inset: rate functions for the same case as main figure at $\tau= 500, 1000, 5000>\tau_{res}$, showing mixing of peaks.} 
\label{fig4}
\end{center}
\end{figure}

All the properties of $P(W_\tau)$ discussed above are retained
when the number of particles is increased. This is shown in Fig.\ref{fignew}, where the probability is plotted for different $N$, at fixed $\phi$, showing an almost-perfect overlap. An $N$-independent distribution means that the surface fraction occupied by the different kind of particles (corresponding to different colours in Fig.\ref{fig1}) is also independent of $N$. Therefore, we argue that the red and blue sectors of the rate functions are mostly originated in the grain boundaries between hexatic domains with different order. Indeed, it has been found~\cite{Dominiesatici} that the size of such hexatic regions does not
scale with the system size, whereas the contribution from the cluster boundary is expected to vanish in the large-$N$ limit.

\subsection{Scaling with the observation time $\tau$}

The above description is entirely based on measurements at fixed $\tau$. We now investigate
the $\tau$ dependence of $I_{\tau}(w)$.  Fig.~\ref{fig3} shows the results for different values of $Pe$ and $\phi=0.5$.
As it is clear, the large deviation principle~(\ref{eq:LDP}) does not hold,
except for $w>w^{\dag}$. Due to the LFT duality discussed previously, 
the corresponding branch of $G_{\tau}(s)$, for $s>s_b$, will also converge
to a well defined limit $G(s)$ as $\tau$ grows large.

For $w<w^\dag$ the suppression of large fluctuations upon increasing $\tau$ is much slower, resulting
in  $\lim_{\tau\rightarrow\infty} I_{\tau}(w)=0$, as it can be seen in Fig.~\ref{fig3}. However,  
our data suggest that $\lim _{\tau \to \infty} \frac{1}{\tau^{\beta}}\ln\left\lbrace Prob.(W_\tau\equiv \tau w)\right\rbrace$ is 
finite for a proper choice of $\beta$. For instance, for $Pe=100$, in the region $w<w^{\dag \dag}$ the best fit yields $\beta \simeq 0.3$. A similar behavior was found in the 
absence of MIPS (see inset of Fig.~\ref{fig2}) in~\cite{cagnetta2017large}. In a shallow region around the minimum at $w=w_c$, a somewhat larger value, compatible with $\beta =0.5$, is
found. These unequal values can be ascribed to the different role played by activity in these two sectors.
On the one hand, a vanishing work ($w=w_c$) is done by particles stacked into a jammed region where the role
of activity is not particularly relevant. On the other hand, a large negative work ($w<w^{\dag \dag}$) is realised by
dense-phase particles pushed against their director, a peculiar phenomenon caused by activity. The inhomogeneous scaling of $P(W_\tau)$'s logarithm with $\tau$ is found not only for different
choices of the P\'eclet number but also for different packing fractions.
The exponent
$\beta $, however, turns out to be parameter dependent. Furthermore, an analogous structure
of the work fluctuations is found for active particles different from colloids, such
as the dumbbells~\cite{Suma14b,Suma14c,petrelli2018active} shown in Fig.1-2 of the SI.

The vanishing of $I(w)$ between the gaseous-phase average $w^{\dag}$ and the typical dense-phase value $w_c=0$ can be physically motivated by considering the probability for a generic particle 
to spend a time $t$ in one of the system phases, which are plotted in 
Fig.~\ref{fig4}. One sees that such
probabilities decay as exponentials, with typical times $\tau _{res}\simeq 300$ or
$\tau _{res}\simeq 100$ for the condensed and
gaseous phase, respectively. Hence, for $\tau \gg  \tau_{res}$ each particle has moved
between the phases several times. Then, considering values of $\tau$ much
larger than those shown in Fig.~\ref{fig3}, namely $\tau \gg \tau _{res}$,
one expects both the double-peaked structure of $I_\tau (w)$ and its
scaling properties to change. Indeed, a smearing of the double-peaked form with $\tau$ can be observed in the upper row of Fig.~\ref{fig3} as well as in inset of Fig.~\ref{fig4}.

A possible explanation for the linear tail at $w<w^{\dag \dag}$, together with its peculiar scaling with $\tau$,
is the following. As already said, disks producing a negative work (blue in Fig.~\ref{fig1}) move along the fluxes
of red particles on the boundaries of hexatically ordered regions.  
Since these channels have a finite size this can occur only up to a certain time $\hat t$, after which the blue particle will be stopped by scattering with stacked cluster molecules. Assuming that such scatterings are independent events the intervals between them is expected to be exponentially distributed as $P(\hat t)\sim e^{-\lambda \hat t}$. The work done in this situation is of order $W_\tau \sim \hat t \, \overline w$, where $\overline w=\tau^{-1} \overline {\bold F_a \cdot \bold s(\tau)}$ is the typical power exerted by the active force in a time $\tau$ when a particle moves a
distance $\bold s$. Hence one has $P(W_\tau)\sim e^{-\lambda W_\tau /\overline w}$. Since the motion is of advective-diffusive nature we argue that $s(\tau)\sim \tau ^{1-\beta}$, where $0<\beta <1$ is some exponent which depends on the diffusive mechanism at work. Therefore $ \overline w\sim -F_{act} \tau^{-\beta}$ and we arrive at $-(1/\tau ^\beta)\ln P(W_\tau)\sim W_\tau/F_{act}$, which agrees with what observed.

\section{Conclusions}

To sum up, we have studied the large fluctuations of the work done by a tagged particle in a system of self-propelled disks, with parameters set so as to have motility-induced phase separation in steady state. With respect to approaches based on collective variables, such as in~\cite{NemotoCates2019}, we found our single-particle approach particularly suitable when two or more phases coexist in the system: on the one hand, it reflects the presence of a dense and a gaseous
phase via a bimodal structure of the probability distribution; on the other hand, it provides informations on additional features of such phases not observed in passive systems. In the homogenoeus state, for instance, one learns about the importance of long-lived micro-clusters~\cite{cagnetta2017large} for the dynamics. In the phase separated state, instead, fluctuations highlight the continuous rearragement of the macroscopic cluster, relevant for phenomena such as cage-breaking and fluidisation of the active dense phase~\cite{mandal2016active}.

Interestingly, the double peaked probability distribution of the active work in the MIPS phase is endowed with two distinct non-convex branches. Correspondingly, the Legendre-Fenchel transform displays a double singularity, differently from what commonly found for instance in liquid-vapor first order transitions. 
For very large observation times $\tau$ we find the bimodal structure to disappear.  The physical reason, as discussed, 
is due to the finite residence time $\tau_{res}$ in both phases. Thus, for $\tau \gg \tau_{res}$ each SPP will have been in and out of the cluster many times. This leads to an homogeneisation of fluctuations, whose general structure does not
reflect anymore the presence of different phases in the system.

This whole pattern of fluctuations is shown to be rather general, being observed also in other models of active brownian particles, as in a system of dumbbells, and can be considered as an hallmark of an active segregation phenomenon.

{\bf Acknowledgements.} F. Cagnetta acknowledges support from the Scottish Funding Council under a studentship.
F.Corberi acknowledges funding from PRIN 2015K7KK8L. 
G.Gonnella acknowledges funding from PRIN 2017/WZFTZP.

\bibliographystyle{apsrev4-1.bst}
\bibliography{dumbbells-biblio}

\end{document}